\documentclass[reprint,superscriptaddress,amsmath,amssymb,aps,prl]{revtex4-2}

\usepackage[T1]{fontenc}  
\usepackage{graphicx}
\usepackage{xcolor}

\begin{document}

\title{\textbf{Colossal Terahertz Magnetoresistance from Magnetic Polarons in EuZn$_2$P$_2$} 
} 

\author{E. Marulanda}
\affiliation{Instituto de Física, Universidade de São Paulo, 05508-090, São Paulo, SP, Brazil}

\author{M. Dutra}
\affiliation{CCNH, Universidade Federal do ABC, 09210-580, Santo André, SP, Brazil}

\author{N. M. Kawahala}
\affiliation{Instituto de Física, Universidade de São Paulo, 05508-090, São Paulo, SP, Brazil}

\author{E. D. Stefanato}
\affiliation{Instituto de Física, Universidade de São Paulo, 05508-090, São Paulo, SP, Brazil}

\author{G. G. Vasques}
\affiliation{CCNH, Universidade Federal do ABC, 09210-580, Santo André, SP, Brazil}

\author{J. Munevar}
\affiliation{Departamento de Física, Universidad del Valle, A. A. 25360, Cali, Colombia}

\author{M. A. Avila}
\affiliation{CCNH, Universidade Federal do ABC, 09210-580, Santo André, SP, Brazil}

\author{F. G. G. Hernandez}
\email{Corresponding author: felixggh@if.usp.br}
\affiliation{Instituto de Física, Universidade de São Paulo, 05508-090, São Paulo, SP, Brazil}

\date{\today}

\begin{abstract}
Magnetic polarons can generate colossal magnetoresistance in magnetic semiconductors, yet their terahertz electrodynamics remain largely unexplored. Here we report magneto-terahertz spectroscopy of the Eu-based Zintl antiferromagnet EuZn$_2$P$_2$. The low-frequency conductivity shows pronounced non-Drude behavior consistent with an evolution from isolated to overlapping magnetic polarons upon cooling. The polaron relaxation time reaches a maximum near the Néel temperature at zero field and exhibits a strong magnetic-field dependence. This polaron-driven reshaping of the conductivity leads to a strongly frequency-dependent magnetoresistance that becomes colossal in the terahertz range, reaching about 90~\% at 1.5~THz, roughly three times larger than the zero-frequency limit value. These results demonstrate that magnetic polarons strongly govern the low-energy electrodynamics and highlight the sensitivity of terahertz spectroscopy to polaronic magnetotransport in correlated magnetic semiconductors.
\end{abstract}


\maketitle
 
\noindent\textit{Introduction}\,---\,Eu-based Zintl phases have emerged as a platform where strong local moments can coexist with low carrier density and narrow gaps, enabling colossal magnetoresistance (CMR) without the mixed valence and perovskite chemistry characteristic of manganites~\cite{Rosa2020Eu5In2Sb6,Wang2021EuCd2P2,Chen2024-gg,PhysRevB.108.205140,llq5-zddh,PhysRevB.111.054406,PhysRevMaterials.8.114421}. In several members of this family, CMR has been linked to magnetic polarons---charge carriers dressed by ferromagnetically aligned spin clouds whose growth and overlap on cooling or under externally applied magnetic fields strongly renormalize charge transport~\cite{Rosa2020Eu5In2Sb6,Souza2022Eu5In2Sb6ESR,Wang2021EuCd2P2,Kopp2026EuCd2P2npj,PhysRevLett.120.257201}. A key open question is how magnetic-polaron transport manifests in the terahertz electrodynamics, where conductivity probes carrier dynamics on picosecond timescales and can reveal relaxation processes and bound states inaccessible to direct current (dc) measurements~\cite{Ulbricht2011RMP,Homes2023EuCd2P2Optical}.

EuZn$_2$P$_2$ is a trigonal (CaAl$_2$Si$_2$-type) Zintl antiferromagnet with A-type order below the Néel temperature $T_\mathrm{N}\approx23.5~\mathrm{K}$~\cite{j1jz-5p73}. Its resistivity shows an insulating-like upturn on cooling~\cite{Berry2022EuZn2P2PRB} and a strong negative magnetoresistance (MR) that reduces the low-temperature resistivity by several orders of magnitude~\cite{Krebber2023EuZn2P2PRB}. Recent transport, electron spin resonance (ESR), and thermodynamic measurements support a magnetic-polaron scenario~\cite{Cook2025EuZn2P2Polaron}. However, while EuZn$_2$P$_2$ has been extensively studied by dc transport and optical spectroscopy~\cite{Krebber2023EuZn2P2PRB,Cook2025EuZn2P2Polaron}, magneto-terahertz measurements on pristine crystals remain largely unexplored. Terahertz time-domain spectroscopy (THz-TDS)~\cite{Baydin2021-je,Koch2023-dp} provides contact-free access to the complex conductivity in the spectral window where itinerant carriers and low-energy excitations compete~\cite{Stefanato2025-hl,RevModPhys.83.471}. Our recent study of Ga-substituted EuZn$_2$P$_2$ at room temperature revealed a pronounced enhancement of the low-frequency electronic response and an increased scattering time, demonstrating the sensitivity of THz-TDS to changes in carrier density and momentum relaxation in this material~\cite{Dutra2025GaEuZn2P2}.

In this Letter we present a magneto-terahertz spectroscopy study of pristine single crystals of EuZn$_2$P$_2$ with zero-field measurements spanning 1.6--150~K and field-dependent measurements up to 7~T at selected temperatures. By parameterizing the low-frequency conductivity we extract its characteristic scales and effective relaxation time, revealing successive regimes consistent with the evolution from isolated to overlapping magnetic polarons and their persistence below $T_\mathrm{N}$. We further quantify the frequency dependence of the MR and show that it becomes colossal at terahertz frequencies, remaining strongly enhanced relative to the zero-frequency limit of the response even above $T_\mathrm{N}$. These results establish terahertz electrodynamics as a powerful probe of magnetic-polaron transport, providing a direct spectroscopic route to identify their coexistence and dynamics and to uncover colossal magnetoresistance in correlated magnetic semiconductors.

\

\noindent\textit{Results}\,---\,We measure the complex terahertz transmission of EuZn$_2$P$_2$ using THz-TDS over the frequency range $\nu=0.2$--2.0~THz. Zero-field spectra span $T=1.6$--150~K, while field-dependent spectra were acquired up to $B=7$~T at selected temperatures. The corresponding complex optical conductivity $\sigma(\nu)$ is obtained as described in the Supplemental Material~\cite{SM}. In the following, we focus on the dissipative (real) component $\sigma_1(\nu)$, which captures the low-energy electronic response relevant to magnetic polarons. 

\begin{figure}
  \centering
  \includegraphics{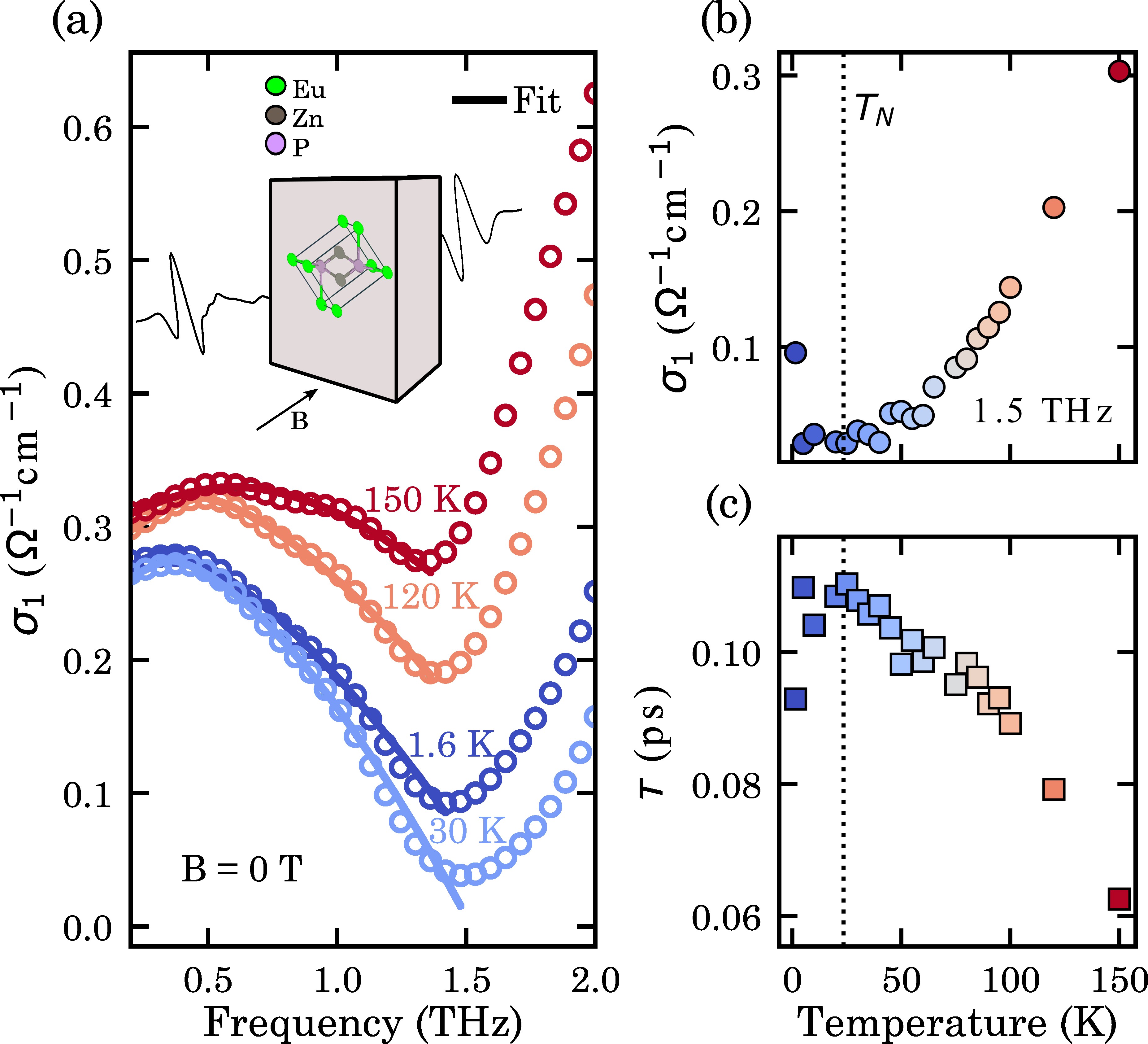}
  \caption{{\bf (a)} Real part of the optical conductivity at selected temperatures in zero magnetic field. Solid lines are power-law fits to the low-frequency response below the conductivity minimum. {\it Inset:} CaAl$_2$Si$_2$-type crystal structure and schematic of the magneto-terahertz time-domain spectroscopy in Faraday geometry. {\bf (b)} Temperature dependence of $\sigma_1$ at 1.5~THz, a representative frequency near the conductivity minimum. {\bf (c)} Effective scattering time extracted from the fits in (a). The dashed line marks the Néel temperature. Error bars are smaller than the data point symbols.}
  \label{fig:fig1}
\end{figure}

We first present zero-field spectra spanning the temperature range where magnetic-polaron phenomenology has been established in EuZn$_2$P$_2$ by transport and local probes~\cite{Krebber2023EuZn2P2PRB,Cook2025EuZn2P2Polaron}. Figure~\ref{fig:fig1}(a) shows $\sigma_1(\nu)$ at several representative temperatures. The low-energy response appears as a broad electronic feature whose maximum, near 0.6~THz at high temperatures, shifts to lower frequencies upon cooling, as seen by the progressive displacement of the low-frequency peak in Fig.~\ref{fig:fig1}(a). Its high-frequency side overlaps with the low-frequency tail of a Eu-related phonon mode, producing a pronounced conductivity minimum at $\nu_\mathrm{min}\!\sim\!1.5~\mathrm{THz}$ that marks the boundary between the two contributions. The phonon peak near 3~THz~\cite{Krebber2023EuZn2P2PRB,PhysRevB.110.014421,Dutra2025GaEuZn2P2} lies outside the usable spectral range of the present experiment, preventing a reliable quantitative determination of its frequency, linewidth, and temperature evolution~\cite{SM}. Nevertheless, the conductivity minimum remains directly resolved and indicates the interplay between the electronic and phonon responses, which has been shown to influence transport in Eu-based compounds hosting magnetic polarons~\cite{Yu2005}.

To isolate the low-energy electronic contribution, we restrict our analysis to frequencies below $\nu_\mathrm{min}$, and parameterize $\sigma_1(\nu)$ with a phenomenological extension of the universal dielectric response, $\sigma_1(\nu)=\sigma_{\rm p} + A\,|\nu-\nu_0|^{2q}$, in which $\nu_0$ allows the conventional power-law form~\cite{Jonscher1977-im, Elliott1987-hf,Tay:23} to represent a finite-frequency maximum. Here $\sigma_{\rm p}$ sets the peak conductivity, while $A$ and $q$ describe the non-Drude curvature of the electronic response. Qualitatively, polaron nucleation and self-trapping should suppress $\sigma_\mathrm{p}$ and shift $\nu_0$ toward lower frequencies, while subsequent growth and overlap recover the spectral amplitude. The concomitant evolution of $A$ and $q$ describes the reshaping of the non-Drude response across these crossovers. Over the measured temperature range, we obtain $A<0$, $0.7\le q\le1$, and $\nu_0$ in the range 0.3--0.6~THz (see Supplemental Material \cite{SM}).

The fitted offset frequency $\nu_0$ corresponds to the maximum of the electronic feature. Its sub-terahertz scale is far below any established electronic energy scale in EuZn$_2$P$_2$. Infrared reflectivity places the indirect and direct gaps at approximately 22 and 80~THz, respectively~\cite{Krebber2023EuZn2P2PRB}. Transport measurements yield activation energies that vary with sample and fit window but consistently fall in the 5--42~THz range, with earlier reports extending to 48--73~THz~\cite{Cook2025EuZn2P2Polaron,Chen2024-gg}. Even the magnetic-field-induced gap renormalization, from the A-type antiferromagnetic to the fully polarized configuration, is on the order of $10^2$~THz~\cite{Cook2025EuZn2P2Polaron}. We therefore interpret $\nu_0$ as a genuine low-energy electronic relaxation scale associated with a dynamical process rather than a conventional electronic excitation, whose origin we relate below to magnetic-polaron dynamics~\cite{mondalInhomogeneousTunnelingNonmonotonic2025}.

Figure~\ref{fig:fig1}(b) shows the temperature dependence of $\sigma_1$ at $\nu=1.5~\mathrm{THz}$, extracted directly from the measured spectra. Because this frequency lies close to the conductivity minimum in Fig.~\ref{fig:fig1}(a), it provides a convenient probe of the high-frequency edge of the electronic feature and its temperature-driven reshaping. Upon cooling, $\sigma_1(\nu\!=\!1.5~\mathrm{THz})$ decreases exponentially from roughly 0.3~$\Omega^{-1}\, {\rm cm}^{-1}$ at 150~K to nearly zero at lower temperatures, consistent with thermally activated behavior. The Arrhenius plot yields an activation energy of $3.5\pm0.3~\mathrm{THz}$ for $T>60~\mathrm{K}$ (see Supplemental Material \cite{SM}), a temperature below which magnetic-polaron correlations have been reported in EuZn$_2$P$_2$~\cite{j1jz-5p73}. 
As observed in EuCd$_2$P$_2$~\cite{Homes2023EuCd2P2Optical}, the Eu-related phonon response exhibits only weak temperature dependence in the complex optical response of closely related Eu-based compounds. Together with the activated behavior and its correlation with the temperature range of magnetic-polaron formation, this suggests that the temperature evolution of the conductivity minimum in Fig.~\ref{fig:fig1}(a) has a substantial contribution from changes in the electronic response associated with magnetic-polaron dynamics, without excluding an additional contribution from the low-frequency phonon tail.

To characterize the relaxation dynamics of the electronic response, we estimate an effective scattering time $\tau$ from the spectral width of this feature. Figure~\ref{fig:fig1}(c) shows $\tau(T)$, extracted from the fitted spectra, with values in the sub-picosecond range. On cooling from 150~K to $\sim\!100~\mathrm{K}$, $\tau$ increases, consistent with a progressive reduction of phonon scattering in semiconductors. Between $\sim\!100~\mathrm{K}$ and $T_\mathrm{N}$, $\tau$ continues to increase but with a slope reduced to roughly one-half of its high-temperature value. This regime coincides with the temperature range where the exponent $q$, which controls the non-Drude curvature of the conductivity, begins to increase, reaching a maximum near 60~K before decreasing again toward $T_\mathrm{N}$ (see Supplemental Material~\cite{SM}). Near $T_\mathrm{N}$, $\tau$ reaches a pronounced maximum and decreases again at lower temperatures. The maximum of $\tau$ near $T_\mathrm{N}$ is consistent with the strong influence of magnetic correlations on relaxation processes associated with magnetic-polaron dynamics, which occurs in sub-picosecond timescales.

\begin{figure}
  \centering
  \includegraphics{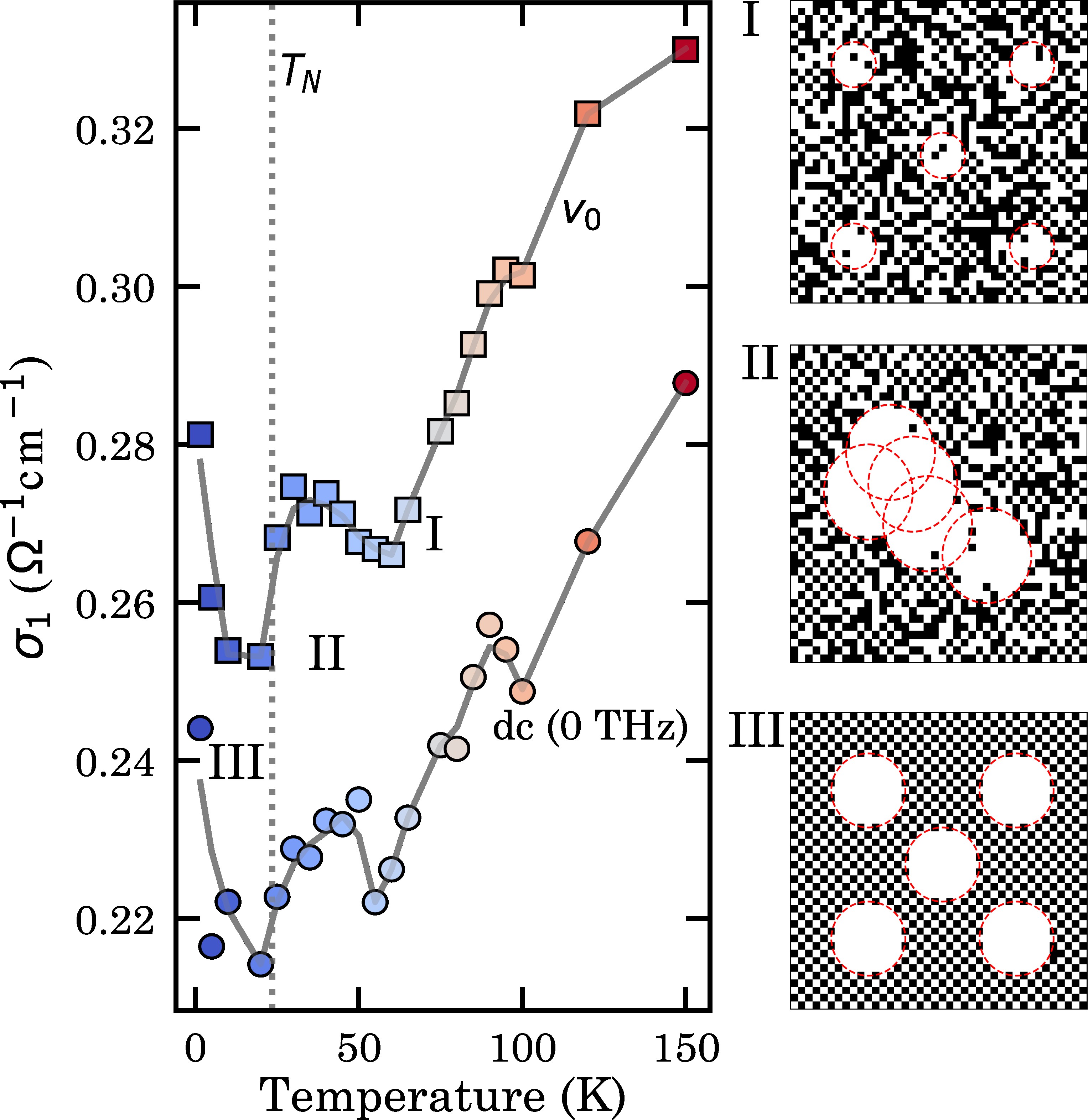}
  \caption{Optical conductivity at dc (circles) and the low-energy scale $\sigma_p$ evaluated at $\nu_0$ (squares), extracted from the fits in Fig.~1. Both datasets are shown on the same vertical scale. Solid lines are guides to the eye. The dashed line marks $T_\mathrm{N}$. {\it Right panels:} schematic spin-plane snapshots (black: down; white: up) illustrating (I) isolated magnetic polarons, (II) their overlap and connection (percolative transport), and (III) the persistence of polaronic correlations in the antiferromagnetic phase. Error bars are smaller than the data point symbols.}
  \label{fig:fig2}
\end{figure}

Figure~\ref{fig:fig2} shows two conductivity scales extracted from the fits in Fig.~\ref{fig:fig1}(a): the dc conductivity $\sigma_{\rm dc}=\sigma_1(0)$ obtained by extrapolating the power law to zero frequency, and $\sigma_{\rm p}=\sigma_1(\nu_0)$, corresponding to the conductivity at the peak of the electronic feature. While $\sigma_{\rm dc}$ reflects the ability of carriers to sustain long-range transport, $\sigma_{\rm p}$ captures the magnitude of the electronic response on terahertz time scales. Both quantities evolve nonmonotonically and exhibit distinct slope changes, indicating successive crossovers in the underlying transport regime.

Within the magnetic-polaron scenario, we distinguish three regimes (I--III) in Fig.~\ref{fig:fig2}. In regime~I ($T\sim100$--60~K), both conductivity scales decrease upon cooling after a subtle increase near $\sim\!100~\mathrm{K}$, with $\sigma_\mathrm{dc}$ remaining systematically smaller than $\sigma_\mathrm{p}$, consistent with carrier self-trapping into ferromagnetic clusters. The onset of this regime coincides with the temperature at which $\tau(T)$ first changes slope, as discussed above, and with ESR evidence for local spin-carrier polarization and magnetic-polaron formation in EuZn$_2$P$_2$~\cite{j1jz-5p73}. In regime~II ($T\sim60~\mathrm{K}$--$T_\mathrm{N}$), both $\sigma_\mathrm{dc}$ and $\sigma_\mathrm{p}$ initially increase upon cooling, reach a maximum in this interval, and decrease again as $T_\mathrm{N}$ is approached, consistent with increasing connectivity of ferromagnetic clusters as they begin to overlap. Finally, in regime~III ($T<T_\mathrm{N}$), both conductivity scales remain finite and continue to evolve, suggesting that ferromagnetic clusters or polaronic correlations persist against the long-range antiferromagnetic background~\cite{PhysRevB.101.205126}. Although CMR is rare in antiferromagnetic materials~\cite{Souza2022Eu5In2Sb6ESR,PhysRevB.57.R8103}, the observed behavior is consistent with reports of ferromagnetic polarons stabilized within antiferromagnetic order in closely related Eu-based materials~\cite{Dawczak-Debicki2024-ry}.

\begin{figure}
  \centering
  \includegraphics{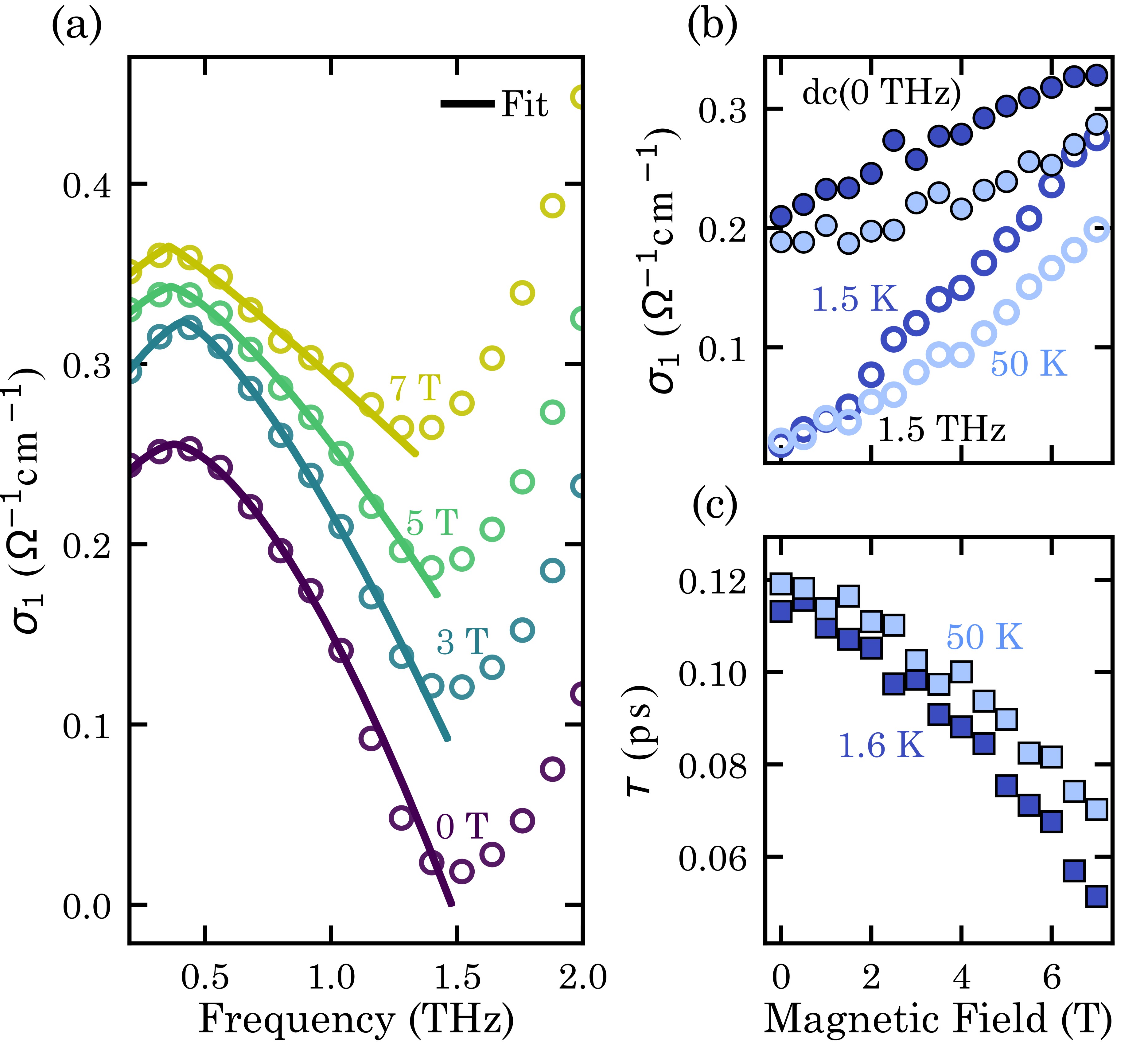}
  \caption{{\bf (a)} Optical conductivity at representative magnetic fields. Solid lines are power-law fits to the low-frequency response. {\bf (b)} Magnetic-field dependence of $\sigma_1$ at 0~THz (dc, filled symbols) and at 1.5~THz (open symbols), measured at $T=1.6$ and 50~K. {\bf (c)} Effective scattering time extracted from the low-frequency fits in (a), shown at the same temperatures as in (b). Error bars are smaller than the data point symbols.}
  \label{fig:fig3}
\end{figure}

We next examine how magnetic fields modify the terahertz conductivity by measuring $\sigma_1(\nu,B)$ at 1.6 and 50 K. The former is the lowest accessible temperature and probes regime III deep in the antiferromagnetic phase, whereas the latter probes the correlated paramagnetic regime II at a comparable absolute temperature offset above $T_\mathrm{N}$. Figure~\ref{fig:fig3}(a) shows $\sigma_1(\nu)$ at 1.6~K for selected fields. As $B$ increases, the conductivity minimum is progressively lifted and the curvature of the electronic feature changes. Although the field dependence of the minimum may also contain a contribution from the low-frequency phonon tail, the concurrent changes in the low-frequency electronic response indicate a substantial electronic contribution to its lifting. We apply the same low-frequency parameterization used in Fig.~\ref{fig:fig1}(a) to the field-dependent spectra. The resulting parameters reveal a systematic field-induced reshaping of the electronic feature. The coefficient $A$ becomes progressively less negative with increasing field at both temperatures. The exponent $q$ starts near 0.8 at zero field but evolves differently at the two temperatures, decreasing toward $\sim\!0.5$ at 1.6~K and increasing to $\sim\!0.9$ at 50~K as $B$ approaches 7~T. By contrast, the offset frequency $\nu_0$ remains within the 0.35--0.5~THz range (see Supplemental Material~\cite{SM}). These trends indicate that the magnetic field primarily modifies the spectral curvature and amplitude of the electronic response, while leaving its characteristic frequency scale largely unchanged.

Figure~\ref{fig:fig3}(b) compares the field dependence of $\sigma_\mathrm{dc}$ and $\sigma_1(\nu\!=\!1.5~\mathrm{THz})$, both of which increase with increasing $B$ at 1.6~K and 50~K. The relative enhancement is systematically larger at 1.5~THz than in $\sigma_\mathrm{dc}$, indicating a stronger field-induced change in the finite-frequency conductivity than in its zero-frequency extrapolation. Within a magnetic-polaron picture, this behavior is expected because the polaron moment couples directly to the magnetic field and modifies its spatial extent, causing polarons aligned with the field to expand while antiparallel ones contract~\cite{PhysRevB.37.4060}. Such field-induced changes in the polaron configuration and overlap can enhance the THz conductivity more rapidly than the long-range transport captured by $\sigma_\mathrm{dc}$. The persistence of this contrast at 50~K is consistent with magnetic-polaron correlations extending well above the onset of long-range order in EuZn$_2$P$_2$~\cite{Krebber2023EuZn2P2PRB}.

Figure~\ref{fig:fig3}(c) shows the effective scattering time $\tau(B)$ extracted from the field-dependent fits. At both 1.6 and 50~K, $\tau$ decreases with increasing $B$, with a stronger suppression observed at low temperature. This trend indicates that the field-induced conductivity enhancement discussed above cannot be attributed to an increase in the relaxation time. Instead, the response is consistent with a field-driven increase in the effective mobile fraction and/or a redistribution of low-energy spectral weight associated with magnetic-polaron correlations. A Drude-equivalent estimate based on the ratio $\sigma_\mathrm{dc}/\tau$ shows that the effective $n/m^*$ scale, where $n$ is the mobile carrier density and $m^*$ the effective mass, increases between 0 and 7~T by factors of approximately 3.4 at 1.6~K and 2.6 at 50~K (see Supplemental Material). A mechanism based on spectral-weight transfer of small polarons was previously reported for changes in the mid-infrared conductivity of CMR manganites~\cite{Alexandrov_1999,PhysRevB.73.024408}. The persistence of this behavior at 50~K further supports the presence of polaronic ferromagnetic correlations above $T_\mathrm{N}$ in EuZn$_2$P$_2$~\cite{Krebber2023EuZn2P2PRB,Dawczak-Debicki2024-ry}.

\begin{figure}
  \centering
  \includegraphics{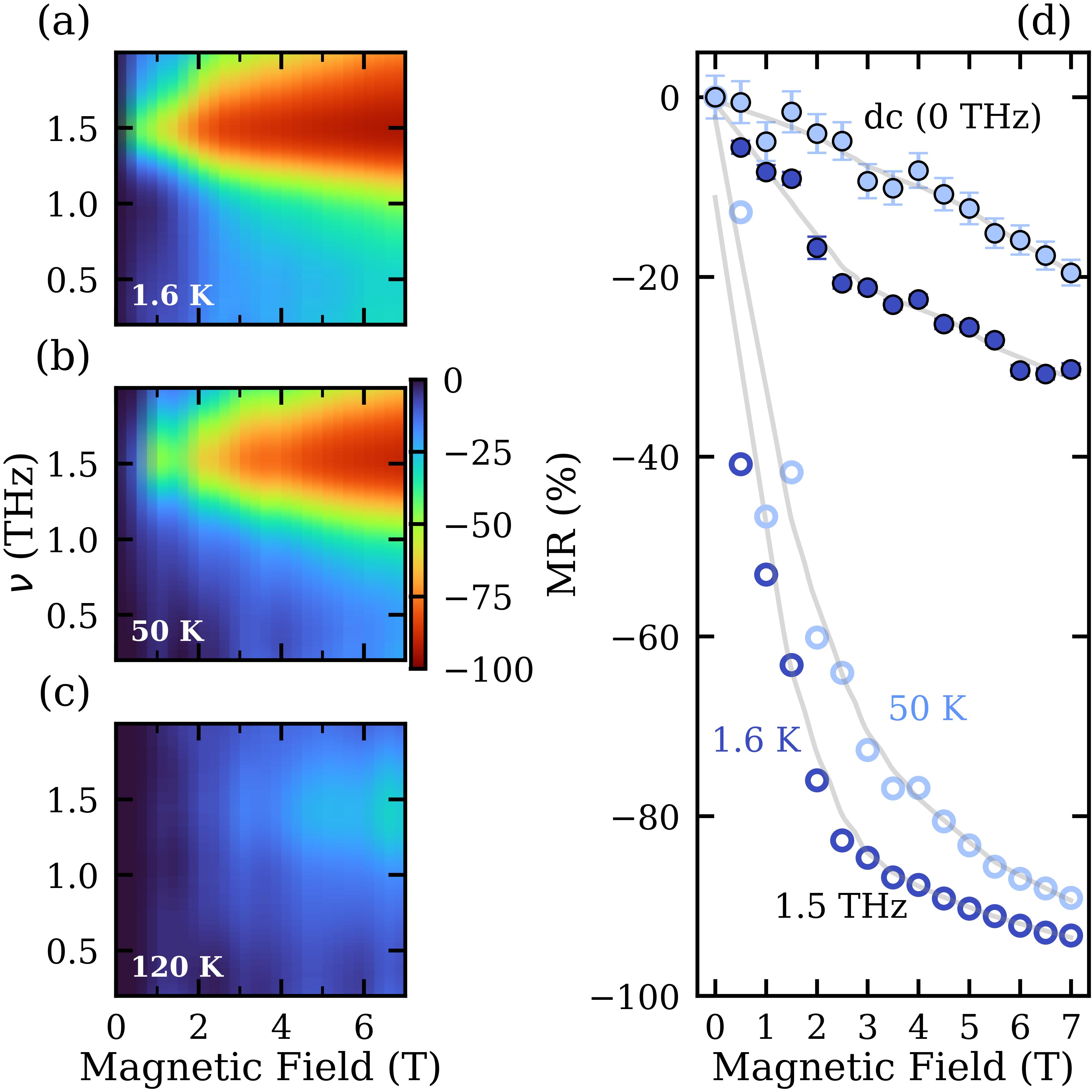}
  \caption{{\bf (a--c)} Terahertz magnetoresistance (\%) as a function of magnetic field and frequency at $T=1.6$, 50 and 120~K. {\bf (d)} Field dependence of the terahertz magnetoresistance at dc (filled symbols) and at 1.5~THz (open symbols), for $T=1.6$~K (dark blue) and 50~K (light blue). Solid lines are guides to the eye. Error bars are shown where visible; otherwise, they are smaller than the data point symbols.}    
  \label{fig:fig4}
\end{figure}

The strong field sensitivity of the conductivity within the magnetic-polaron scenario suggests that the magnetoresistance should exhibit a pronounced frequency dependence in the terahertz range. To quantify this dynamical counterpart of the dc magnetoresistance, we define a conductivity-based terahertz magnetoresistance $\mathrm{MR}(\nu,B)=\left[\sigma_1(\nu,0)-\sigma_1(\nu,B) \right]/\sigma_1(\nu,B)$. Figure~\ref{fig:fig4}(a--c) shows MR maps as a function of magnetic field and frequency at 1.6~K, 50~K, and 120~K.

At $T=1.6$~K (Fig.~\ref{fig:fig4}a), negative MR exhibits a pronounced cone-like structure centered near $\nu\!\sim\!1.5~\mathrm{THz}$, reaching nearly 100~\% in the vicinity of the conductivity minimum at high fields. The enhancement appears already at low magnetic fields and broadens progressively with increasing $B$, bounded by a 50~\% MR contour that expands to roughly 1~THz at 7~T. Outside this region the response remains comparatively weak, staying near zero for $B<2$~T and reaching only $\sim\!30\%$ at higher fields. A similar pattern persists at 50~K (Fig.~\ref{fig:fig4}b), although with reduced overall intensity, indicating that the terahertz MR remains substantial even above $T_\mathrm{N}$. By contrast, at 120~K (Fig.~\ref{fig:fig4}c) the map is dominated by weak responses, with only a faint remnant of the cone-like structure visible and maximum values below $\sim\!30\%$.

To facilitate comparison between the terahertz response and its zero-frequency extrapolation, Fig.~\ref{fig:fig4}(d) shows MR curves extracted from the maps at $\nu=1.5~\mathrm{THz}$ together with the extrapolated dc MR. At both 1.6~K and 50~K the terahertz MR rises steeply at low fields and gradually approaches a saturation-like behavior at intermediate fields. At 7~T, $\mathrm{MR}(\nu\!=\!1.5~\mathrm{THz})$ reaches $\sim\!93~\%$ at 1.6~K and $\sim\!89~\%$ at 50~K, whereas the dc MR remains near $\sim\!30~\%$ and $\sim\!20~\%$, respectively. This large disparity between terahertz and dc MR establishes a colossal terahertz magnetoresistance in EuZn$_2$P$_2$, demonstrating that terahertz conductivity is exceptionally sensitive to field-induced changes in the electronic response associated with magnetic-polaron correlations.

\

\noindent\textit{Discussion}\,---\,The observation of a colossal terahertz magnetoresistance in EuZn$_2$P$_2$ places our results in the broader context of large magnetoresistive phenomena that have attracted intense recent interest because of their fundamental importance and potential for magnetic control of electrical conductivity~\cite{PhysRevB.110.064407,PhysRevB.110.174408,PhysRevB.111.014431,PhysRevB.111.115114,p2c5-r163,6yv6-kf97,wq1l-wj9k,7tkr-16rc}. At terahertz frequencies, large magnetoresistive responses have been reported most prominently in manganites and, more recently, in spin-textured antiferromagnets. In manganites, negative THz MR have been measured in La$_{1-x}$Sr$_x$MnO$_3$, reaching about $40\%$ in 8 T at 1~THz~\cite{LloydHughes2017THzCMR} and nearly $60\%$ in 6~T at 0.4~THz~\cite{Tay:23}, depending on composition, both near the Curie temperature. These effects have been discussed in terms of intrinsic intragrain transport associated with double exchange, hopping between localized states, and phase competition. Scaling relations between conductivity and terahertz dielectric permittivity have also been investigated in the CMR regime of Pr$_{0.65}$Ca$_{0.28}$Sr$_{0.07}$MnO$_3$~\cite{Pimenov2006THzScaling}. 

Similar to our observations, Ref.~\cite{LloydHughes2017THzCMR} reports a THz MR larger than the extrapolated dc value. By contrast, the $\sim\!30\%$ MR inferred here from the zero-frequency extrapolation at 1.6~K and 7~T is far smaller than the direct dc CMR reported for pristine EuZn$_2$P$_2$, which reaches 99.4\% at 4~T and 99.7\% at 9~T near $T_\mathrm{N}$~\cite{Krebber2023EuZn2P2PRB,Cook2025EuZn2P2Polaron}. This discrepancy can be understood in terms of the different time and length scales probed by these techniques. While dc transport is sensitive to long-range conduction paths, terahertz spectroscopy probes local carrier dynamics and is therefore less affected by inhomogeneous transport or slow channels lying below the experimental frequency window. Additional differences may also arise from extrinsic effects such as electrical contacts.

More recently, giant negative THz MR in EuTe$_2$ was linked to a distinct mechanism based on magnetic-field control of spin orientation and the associated reconstruction of the low-energy band structure, including a field-induced gap change. In that system, MR exceeding 90~\% was reported in 2.5~T at 1.1~THz for temperatures slightly above $T_\mathrm{N}$~\cite{Hao2026EuTe2THzMR}. In comparison, the terahertz MR observed here in EuZn$_2$P$_2$ reaches the same record level above 4~T at 1.6~K, albeit at higher frequency, placing it among the largest responses reported to date. 
Importantly, the mechanism responsible for the effect is fundamentally different from those previously reported. Our results indicate a strong magnetoresistive response governed by magnetic polarons, demonstrating that polaronic magnetotransport remains active in terahertz frequencies for this class of correlated magnetic semiconductors.

\

\noindent\textit{Conclusions}\,---\,In summary, our time-domain magnetospectroscopy measurements reveal that magnetic polarons strongly reshape the terahertz optical conductivity of the Eu-based Zintl antiferromagnet EuZn$_2$P$_2$, producing a colossal negative terahertz magnetoresistance that exceeds 90~\% at 1.5~THz below the magnetic transition temperature. The magnetoresistance at terahertz frequencies is roughly three times larger than the value inferred in the zero-frequency limit and becomes large already at moderate magnetic fields, demonstrating a pronounced frequency-dependent magnetotransport response. At the same time, the low-energy electronic response associated with magnetic polarons exhibits a maximum lifetime near the Néel temperature at zero field, highlighting the strong coupling between spin correlations and charge dynamics in this system. These results establish magnetic polarons as a key mechanism governing the terahertz electrodynamics of this class of correlated magnetic semiconductors. The substantial magnetic control of the polaronic terahertz conductivity observed here may also be of interest for magnetically tunable terahertz devices.

\

\noindent\textit{Acknowledgements}\,---\,We acknowledge financial support from the São Paulo Research Foundation (FAPESP) under Grant Nos. 2021/12470-8, 2023/04245-0, and 2023/16742-8, and by the National Council for Scientific and Technological Development (CNPq) under Grant No. 306550/2023-7. E.M. acknowledges financial support from Grant No. 88887.007580/2024-00 of the Coordenação de Aperfeiçoamento de Pessoal de Nível Superior (CAPES). N.M.K. acknowledges support from FAPESP Grant No. 2023/11158-6. E.D.S. acknowledges financial support from CNPq Grant No. 407815/2022-8 and from FAPESP Grant No. 2025/02029-3. Additional support from the INCT Advanced Quantum Materials project funded by CNPq (Grant No. 408766/2024-7), FAPESP (Grant No. 2025/27091-3), and CAPES is also gratefully acknowledged. Activities within the GMQ/UFABC group were supported by FAPESP under Grant No. 2017/10581-1, CNPq under Grant Nos. 309363/2022-5 and 405408/2023-4, CAPES, and CEM/UFABC.

\

\noindent\textit{Data availability}\,---\,The data that support the findings of this article are available from the authors upon reasonable request.

\bibliography{bibl}

\clearpage
\onecolumngrid


\renewcommand{\thesection}{S\arabic{section}}   
\renewcommand{\thetable}{S\arabic{table}}   
\renewcommand{\thefigure}{S\arabic{figure}}
\renewcommand{\theequation}{S\arabic{equation}}

\renewcommand{\figurename}{Fig.}
\renewcommand{\tablename}{Table}

\setcounter{equation}{0}
\setcounter{figure}{0}
\section*{Supplemental  Material}

\section{S1. Samples}

High-purity single crystals of EuZn$_2$P$_2$ were synthesized by the Sn-flux method. 
High-purity elements, Eu (99.9\%), Zn (99.999\%), P (99.999\%), and Sn (99.999\%) from Alfa-Aesar were weighted in an atomic ratio of 1:2:2:40 and placed inside a quartz tube with quartz wool.
The evacuated and sealed ampoule was gradually heated to 500~ºC over 2~h, kept at this temperature for 1~h, then further heated to 1150~ºC over 4~h and held at this temperature for 10~h. 
Controlled cooling was carried out down to 850~ºC at a rate of 2~ºC/h, after which the tubes were rapidly spun to separate the crystals from the flux. More details can be found in references \cite{j1jz-5p73,Dutra2025GaEuZn2P2}.

\section{S2. Experimental Details}

\noindent\textbf{Experimental setup}

\begin{figure}[h!]
    \centering
    \includegraphics[width=0.9\linewidth]{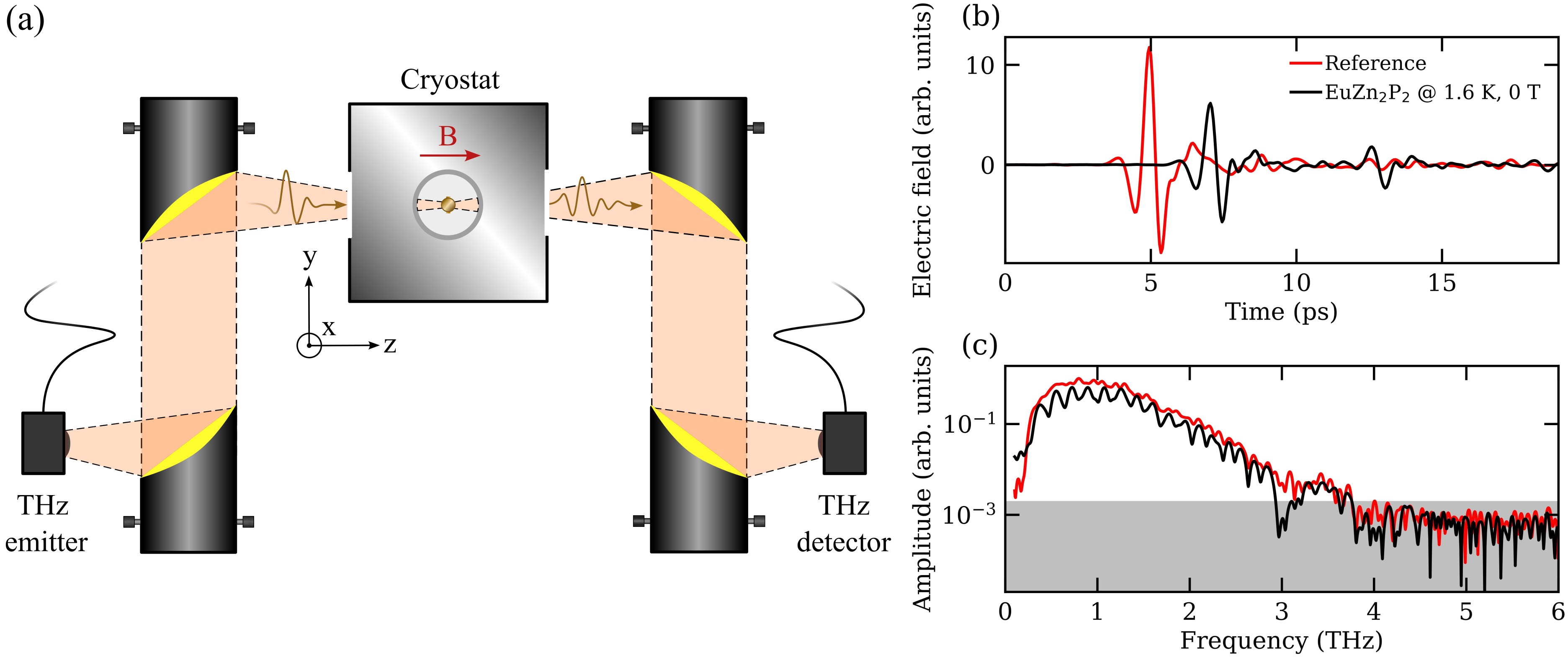}
    \caption{
        (a) Schematic of the free-space coupling between the THz emitter and detector through the cryomagnet, with the sample measured in Faraday geometry. (b) Representative reference and EuZn$_2$P$_2$ time-domain electric fields at 1.6~K and 0~T. (c) Corresponding raw Fourier amplitudes; the shaded region indicates the approximate experimental background level.
    }
    \label{fig:S1}
\end{figure}

Terahertz time-domain spectroscopy measurements were performed in transmission using a commercial fiber-based spectrometer (TeraFlash pro, TOPTICA Photonics). The system employs photoconductive antennas for THz generation and coherent detection, driven by 1560-nm femtosecond laser pulses, with integrated optical delay lines for time-domain acquisition. The spectrometer was externally coupled in free space to a cryogen-free superconducting magnet system (SpectromagPT, Oxford Instruments) equipped with $z$-cut quartz optical windows. Off-axis parabolic mirrors were used to focus the THz beam onto the sample and recollect the transmitted radiation onto the detector. Because its lateral dimensions were substantially smaller than the typical THz beam spot of approximately 3~mm at the sample position, the EuZn$_2$P$_2$ crystal was mounted over a 0.7-mm-diameter aperture. Measurements were performed in Faraday geometry, with the THz propagation direction parallel to the external magnetic field, over the temperature range 1.6--150~K and in fields up to 7~T. Figure~\ref{fig:S1}(a) illustrates the magneto-THz experimental configuration.

\

\noindent\textbf{Data acquisition and conductivity extraction}

Reference and sample time-domain electric fields, $E_{\mathrm{ref}}(t)$ and $E_{\mathrm{sam}}(t)$, were acquired through the same magneto-optical path. The waveforms used in the conductivity analysis were processed using the same time-domain window to suppress delayed Fabry--P\'erot echoes, following the procedure described in Ref.~\cite{Marulanda2025-wb}. Their Fourier transforms were then used to obtain the complex transmission coefficient, $T(\nu)=E_{\mathrm{sam}}(\nu)/E_{\mathrm{ref}}(\nu)$. The complex optical conductivity, $\sigma(\nu)=\sigma_1(\nu)+i\sigma_2(\nu)$, was extracted using the standard Fresnel inversion for a plane-parallel slab~\cite{Lloyd-Hughes2012-es}. We focus the quantitative analysis on the dissipative component $\sigma_1(\nu)$, which is directly constrained by the measured attenuation and contains the conductivity scales and effective relaxation time relevant to magnetic-polaron transport. It is also experimentally more robust than $\sigma_2(\nu)$, which in bulk samples is particularly sensitive to small phase and thickness uncertainties. For completeness, representative $\sigma_2(\nu)$ spectra are presented in Fig.~\ref{fig:S6} and show qualitative behavior consistent with the localized non-Drude response inferred from $\sigma_1(\nu)$.

\

\noindent\textbf{Usable spectral range}

Figures~\ref{fig:S1}(b) and \ref{fig:S1}(c) show representative reference and sample time-domain waveforms and their corresponding raw Fourier amplitudes, respectively. These raw spectra are shown to illustrate the signal available before the time-domain windowing used in the conductivity extraction. Although the bare spectrometer can provide spectral coverage up to approximately 6~THz under optimized conditions, the usable bandwidth of the complete magneto-THz configuration is reduced by propagation through the cryomagnet optical windows, the 0.7-mm sample-holder aperture, and the frequency-dependent transmission of the sample. Residual field amplitude remains detectable above 2~THz, but the sample signal decreases rapidly and approaches the experimental background near the reported phonon frequency around 3~THz. In this regime, the sample-to-reference amplitude ratio and, particularly, the transmission phase are not sufficiently robust for a reliable Fresnel inversion. We therefore restrict the quantitative conductivity analysis to 0.2--2.0~THz, over which the complex transmission is reproducible across temperature and magnetic field.

\

\noindent\textbf{Arrhenius analysis}

\begin{figure}[h!]
  \centering
  \includegraphics{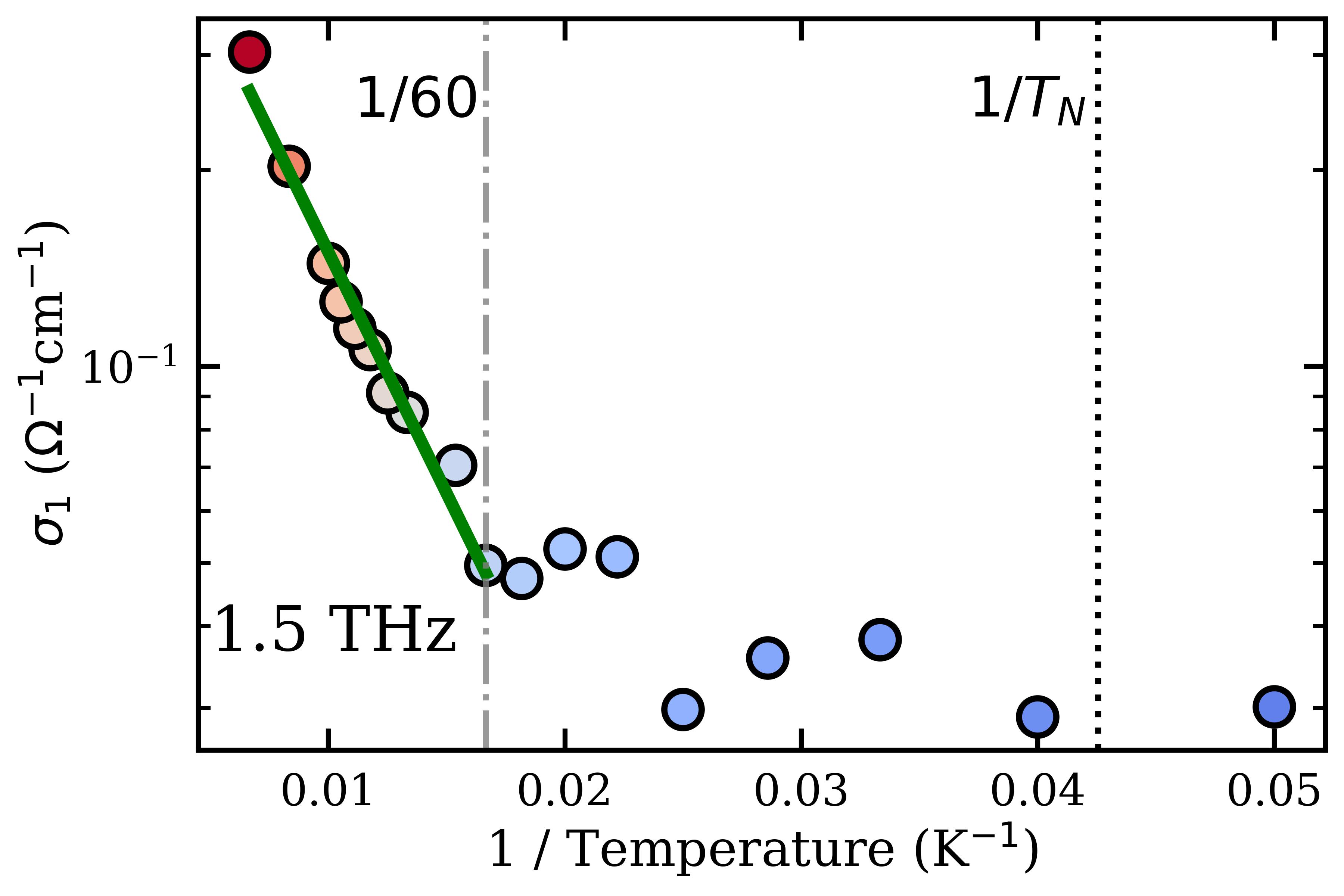}
  \caption{Semilogarithmic plot of the real part of the optical conductivity, $\sigma_1$, at 1.5 THz for $\mathrm{EuZn}_2\mathrm{P}_2$ as a function of inverse temperature. The dashed line is the Arrhenius fit over $60\ \mathrm{K} \leq T \leq 150\ \mathrm{K}$. The dot-dashed and dotted lines mark 60 K and $T_N = 23.5$ K, respectively.}
  \label{fig:S2}
\end{figure}

An Arrhenius analysis of the high-temperature THz conductivity in $\mathrm{EuZn}_2\mathrm{P}_2$ Fig.~\ref{fig:S2} yields an activation energy of $E_a/h = 3.5 \pm 0.3$ THz ($\approx 14.5$ meV). In the semilogarithmic representation of $\sigma_1$ at 1.5 THz versus inverse temperature, the data follow the activated form $\sigma_1(T)\propto \exp(-E_a/k_B T)$ over $60\ \mathrm{K} \leq T \leq 150\ \mathrm{K}$. The deviation from this behavior below $\sim 60$ K occurs well above the Néel temperature, $T_N = 23.5$ K. The extracted activation energy is approximately six times smaller than the indirect semiconductor gap, $E_g/h \approx 22$ THz ($E_g \approx 91$ meV), reported in Ref.~[15], and is therefore too small to be naturally associated with conventional intrinsic carrier activation governed by the semiconductor gap. Instead, it identifies a lower-energy subgap activation channel. Given the independent transport, ESR, and thermodynamic evidence for magnetic-polaron formation in $\mathrm{EuZn}_2\mathrm{P}_2$~[13,16], we discuss this channel within the same polaronic framework. The Arrhenius analysis alone, however, does not establish whether $E_a$ corresponds to a magnetic-polaron binding energy or to a specific in-gap polaron state.

\

\noindent\textbf{Fit parameters}

\begin{figure}[h!]
  \centering
  \includegraphics{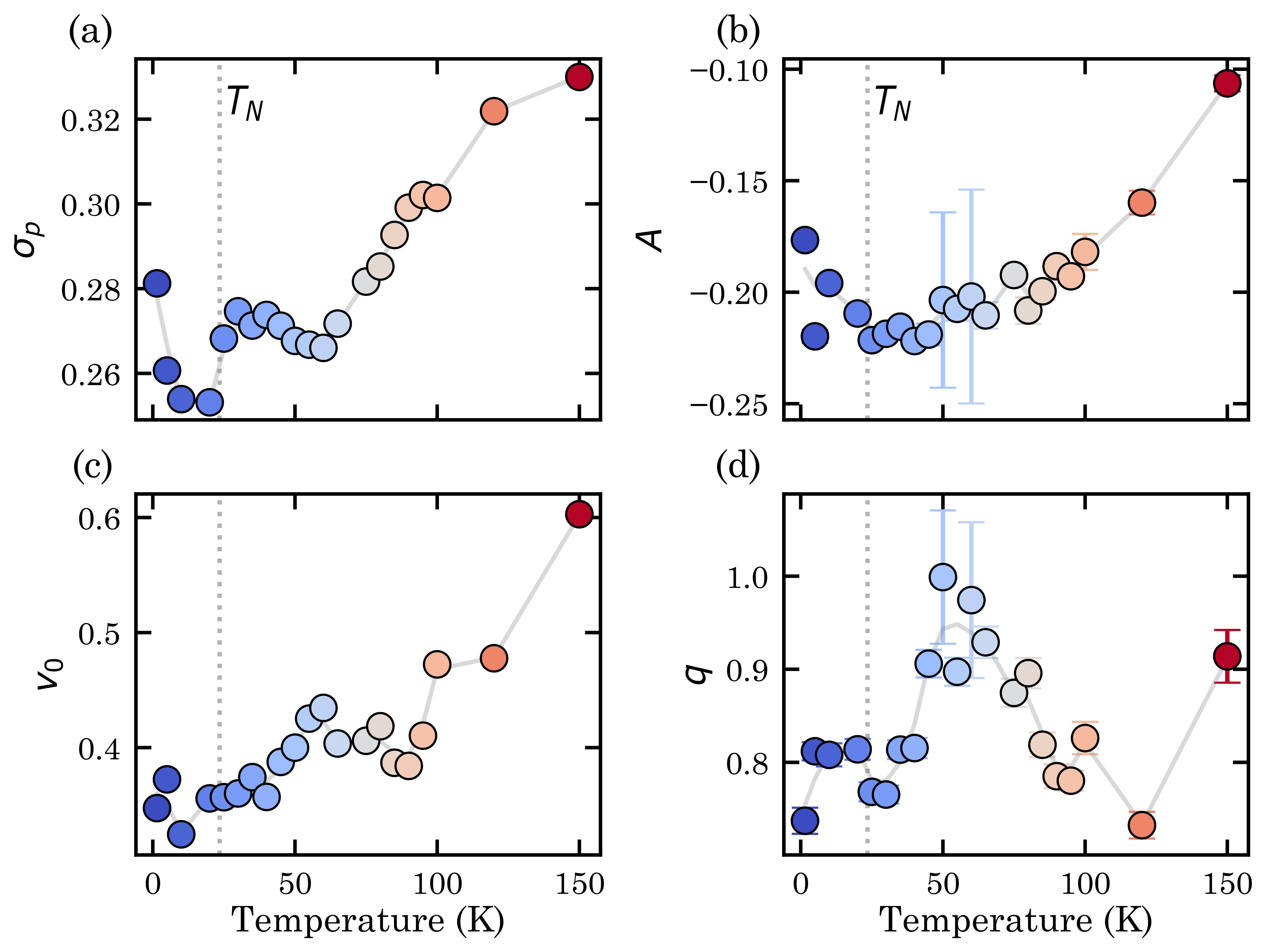}
  \caption{Temperature dependence of the fit parameters:
(a) $\sigma_p$ ($\Omega^{-1}\,\mathrm{cm}^{-1}$), (b) $A$ ($\Omega^{-1}\,\mathrm{cm}^{-1}\,\mathrm{THz}^{-2q}$), (c) $\nu_0$ (THz), and (d) $q$.
The vertical dotted line marks $T_N$.
Solid lines are guides to the eye.}
  \label{fig:S3}
\end{figure}

\begin{figure}[h!]
  \centering
  \includegraphics{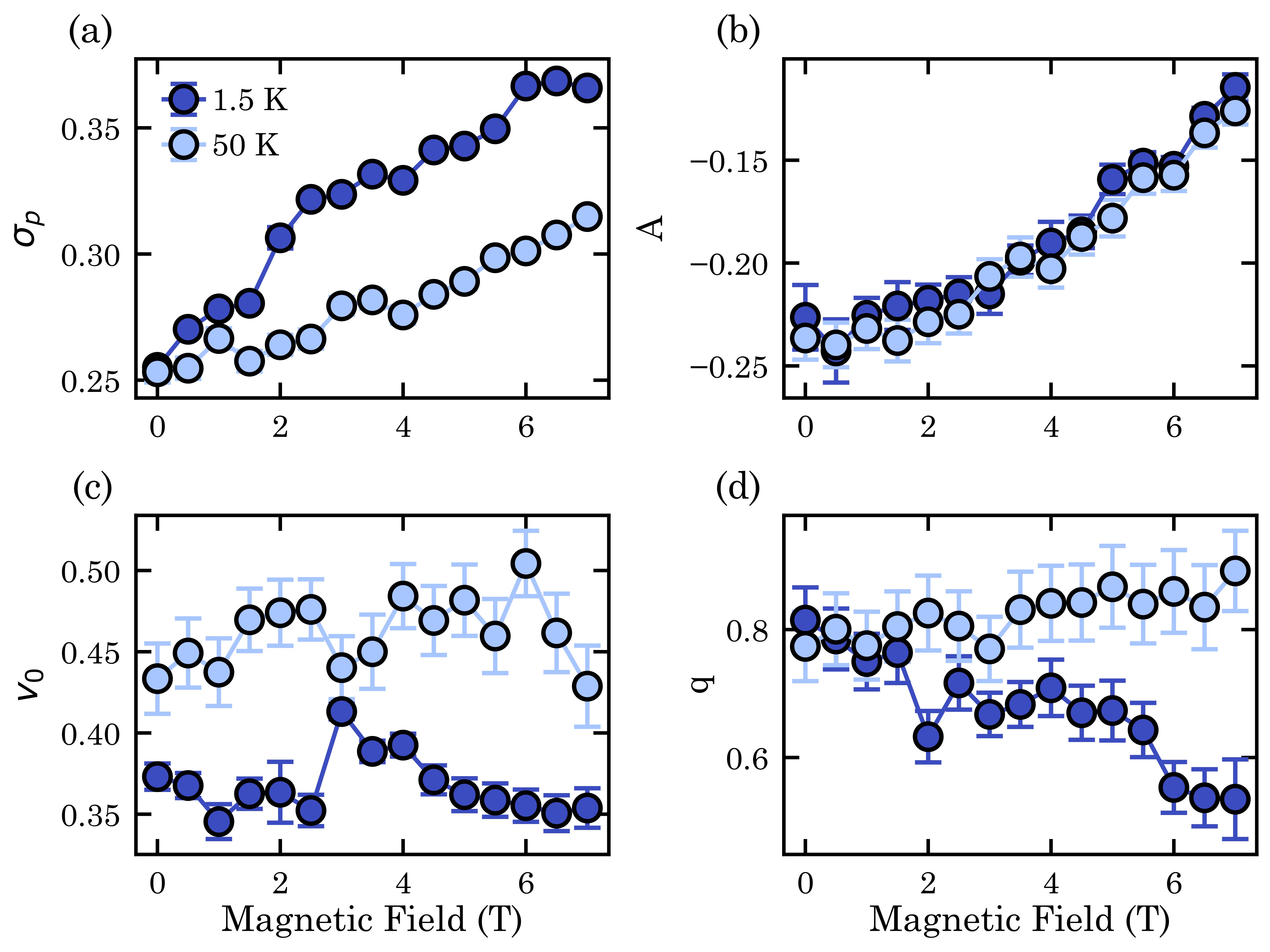}
  \caption{Magnetic-field dependence of the fit parameters:
(a) $\sigma_p$ ($\Omega^{-1}\,\mathrm{cm}^{-1}$), (b) $A$ ($\Omega^{-1}\,\mathrm{cm}^{-1}\,\mathrm{THz}^{-2q}$), (c) $\nu_0$ (THz), and (d) $q$.
Solid lines are guides to the eye.}
  \label{fig:S4}
\end{figure}

The fit parameters from $\sigma(\nu)=\sigma_p + A(\nu-\nu_0)^{2q}$ are shown as a function of temperature and magnetic field in Figs.~\ref{fig:S3} and \ref{fig:S4}, respectively.

\

\noindent\textbf{Field-induced effective $n/m^*$ scale}

\begin{figure}[h!]
    \centering
    \includegraphics{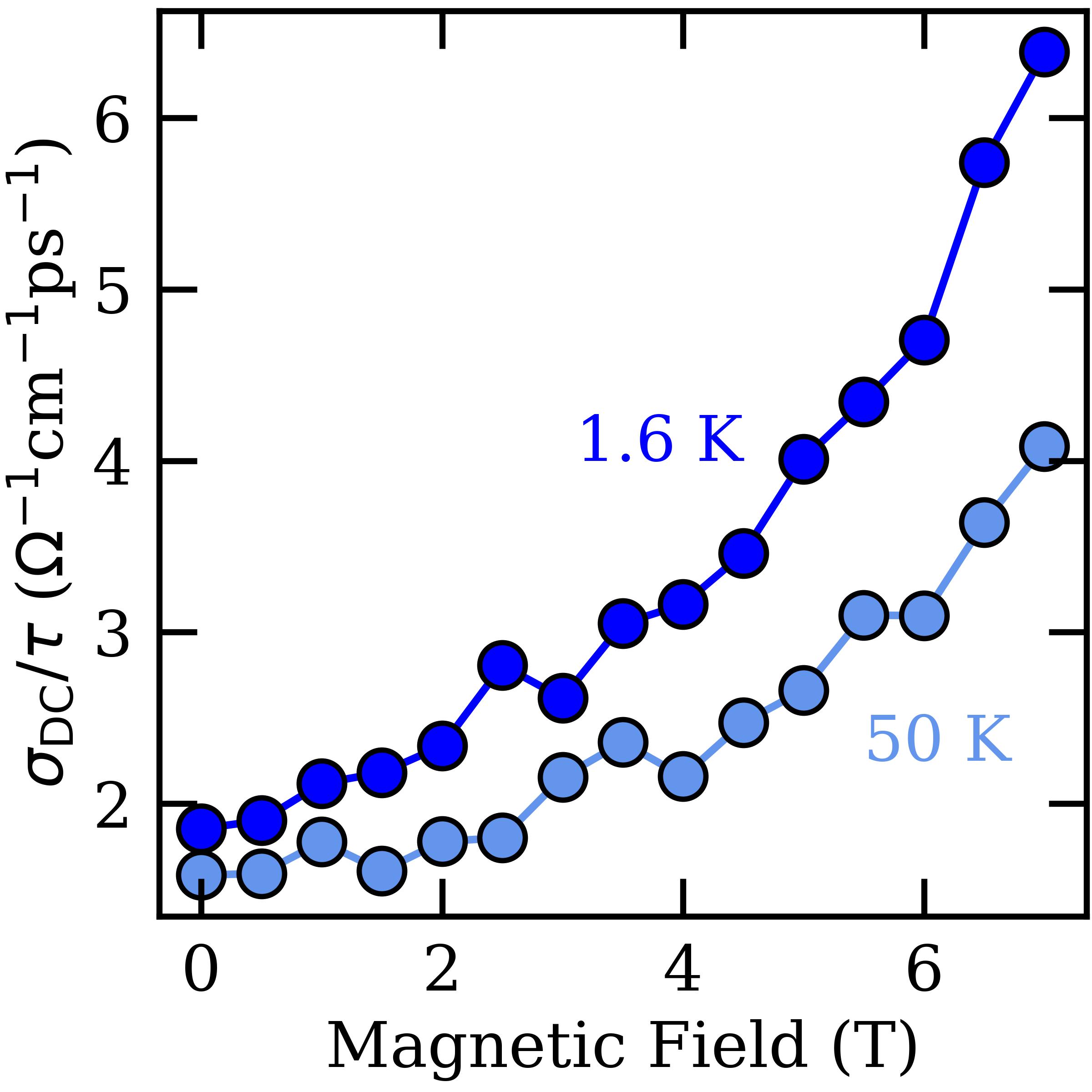}
    \caption{Magnetic-field dependence of the Drude-equivalent ratio $\sigma_\mathrm{dc}/\tau$ at 1.6 and 50 K. Solid lines are guides to the eye.}
    \label{fig:S5}
\end{figure}

\begin{figure}[h!]
    \centering
    \includegraphics{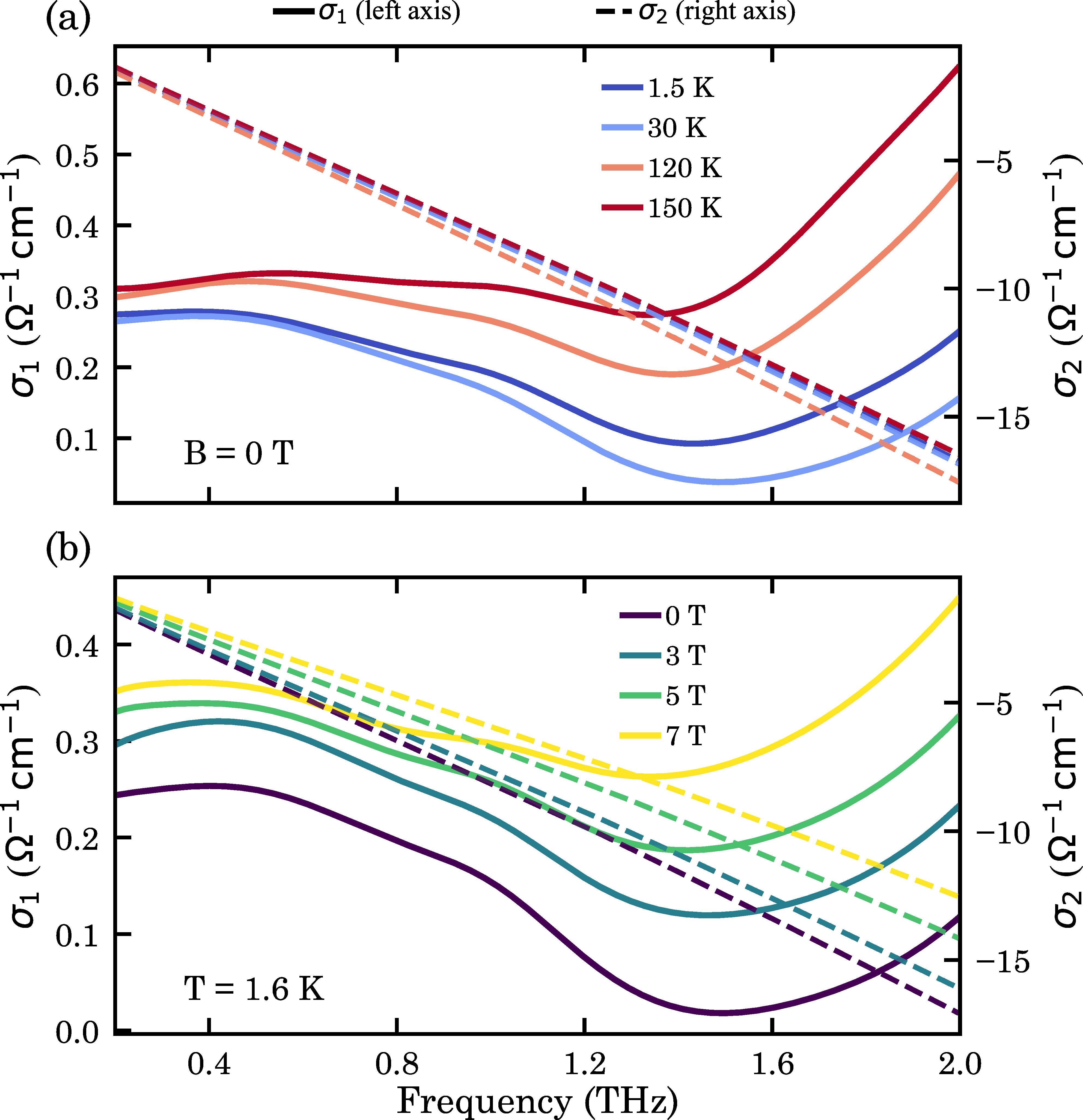}
    \caption{Representative real ($\sigma_1$, solid lines) and imaginary ($\sigma_2$, dashed lines) parts of the optical conductivity of EuZn$_2$P$_2$: (a) temperature dependence at B=0 T and (b) magnetic-field dependence at T=1.6 K.}
    \label{fig:S6}
\end{figure}

To provide a semi-quantitative estimate of the field-induced change in the effective $n/m^*$ scale, we evaluate the Drude-equivalent ratio $\sigma_\mathrm{dc}/\tau$, where $\sigma_\mathrm{dc}$ is the extrapolated zero-frequency conductivity and $\tau$ is the effective relaxation time extracted from the low-frequency fits. In the Drude limit, this ratio is proportional to $ne^2/m^*$. Since the low-energy response of EuZn$_2$P$_2$ is distinctly non-Drude, this relation should be regarded only as a benchmark for the effective $n/m^*$ scale rather than as a direct determination of the carrier or polaron density. As shown in Fig.~\ref{fig:S5}, $\sigma_\mathrm{dc}/\tau$ increases between 0 and 7~T by factors of approximately 3.4 at 1.6~K and 2.6 at 50~K.

\

\noindent\textbf{Imaginary part of the optical conductivity}

For completeness, Fig.~\ref{fig:S6} shows representative spectra of the imaginary part of the EuZn$_2$P$_2$ optical conductivity together with its real counterpart. For all temperatures and magnetic fields shown, $\sigma_2$ remains negative and decreases monotonically with frequency, consistent with localized non-Drude conductivity spectra reported in previous terahertz studies~\cite{phanindraEpitaxialStrainDriven2017}. Its qualitative behavior is consistent with the interpretation inferred from $\sigma_1$. Since the characteristic conductivity scales and effective relaxation time investigated in the present work are more directly and robustly represented by $\sigma_1$, we restrict the quantitative analysis to the real part of the conductivity.

\section{S3. Supplemental Figures}

\end{document}